\documentstyle[12pt]{article}
\begin{document}
\title{A Brief Note on Jupiter's Magnetism}
\author{B.G. Sidharth$^*$\\
Centre for Applicable Mathematics \& Computer Sciences\\
B.M. Birla Science Centre, Adarsh Nagar, Hyderabad - 500 063 (India)}
\date{}
\maketitle
\footnotetext{$^*$Email:birlasc@hd1.vsnl.net.in; birlard@ap.nic.in}
\begin{abstract}
A recent model which gives the contribution of the earth's solid core
to geo magnetism is seen to explain Jupiter's magnetism also.
\end{abstract}
As is known Jupiter exhibits an earth like dipole magnetism $\sim 10^4$ times
that of the earth. Geo magnetism is explained by the dynamo model of the
earth's liquid core. The planet Jupiter however is qualitatively
different. Though like the earth it has a hot solid metallic core, which
has a radius of about 20 times that of the earth's solid core, this is
however surrounded by a 60,000 kilometres thick Hydrogen mantle\cite{r1}.\\
Recently it was suggested that the earth's solid core contributes significantly
to geomagnetism\cite{r2}. This is based on the fact that below the Fermi
temperature, Fermions would have an anomalous semionic character - that is
they would obey a statistics inbetween the Fermi-Dirac and the Bose-Einstein
\cite{r3}. This would have the consequence that the magnetization density
in such a situation, which is given by the well known expression\cite{r4}
\begin{equation}
M = \frac{\mu (2N_+ - N)}{V}\label{e1}
\end{equation}
where $\mu$ is the electron magnetic moment, $N_+$ is the average number
of Fermions with spin up out of an assembly of $N$ Fermions, now has the
value $\sim \frac{\mu N}{V}$, owing to the fact that
\begin{equation}
\frac{1}{2} < \frac{N_+}{N} < 1\label{e2}
\end{equation}
Inequality (\ref{e2}) expresses the semionic behaviour.\\
Remembering that the core density of Jupiter is of the same order as that
of the earth, while the core volume is about $10^4$ times that of the earth,
we have $N \sim 10^{52}$, so that the magnetization $MV$, from (\ref{e1})
$\sim 10^4$ times the earth's magnetism, as required.\\
Finally, as pointed out in ref.\cite{r2}, owing to the semionic behaviour as
expressed by (\ref{e2}), the magnetization would be sensitive to external
magnetic influences, and we could have magnetic reversals as in the case
of the earth.\\
Incidentally it may be pointed out that the same model could also explain
the magnetism of Neutron stars and White Dwarfs (cf.ref.\cite{r2}).

\end{document}